\documentclass[journal=jacsat,manuscript=article]{achemso}

\usepackage[version=3]{mhchem} % Formula subscripts using \ce{}
\usepackage{graphicx}
\usepackage{tikz}
\usepackage{caption}
\usepackage{lineno}
\usepackage{multirow}
\usepackage{url}

\author{Marco Biagi}
\email{marcobiagi97@gmail.com}

\author{Corrado C. M. Capriata}

\author{Corentin Bouchard}
\affiliation{CEA-Leti Minatec, Grenoble, France}

\author{Subham Kintali Senapati}

\affiliation{Univ. Grenoble Alpes, CEA, CNRS, Grenoble-INP, SPINTEC}
\author{Ioannis Trikoilis Koll}

\author{Ricardo C. Sousa}
\affiliation{Univ. Grenoble Alpes, CEA, CNRS, Grenoble-INP, SPINTEC}

\author{Louis Hutin}

\author{Bernard Viala}
\affiliation{CEA-Leti Minatec, Grenoble, France}

\author{Kevin Garello}
\email{kevin.garello@cea.fr}
\affiliation{Univ. Grenoble Alpes, CEA, CNRS, Grenoble-INP, SPINTEC}

\title[]
  {Orbital and Spin-Orbit Torque Interplay in Ta/W-based Magnetic Tunnel Junctions with Vertical Non-local Switching}
\abbreviations{IR,NMR,UV}

\begin{document}
%\linenumbers


%TC:ignore
\begin{abstract}
Spin-orbit torque (SOT) enables ultra-fast, energy-efficient magnetization switching, making it a promising mechanism for introducing MRAMs for cache memory applications. However, current SOT-MRAM devices face write efficiency limitations, with charge-to-spin conversion ($\xi_{DL}$) reaching $\sim$ 45\%, far below the projected $\sim$ 80\% needed to comply with the current delivery of advanced transistor nodes. Recent advances in orbital current physics, evidenced in a wide class of materials, offer a path to enhance $\xi_{DL}$. Here, we study the Ta(3–30 nm)\slash W(1–4 nm) system, revealing a large additional spin-orbit torque contribution arising from Ta, a four-fold increase compared to the spin Hall effect in Ta alone, attributed to the orbital Hall contribution. This system exhibits larger $\xi_{DL}$ than W-based SOT systems with more robust perpendicular magnetic anisotropy and compatibility with 400$^\circ$C annealing. Leveraging these advantages, we integrate the Ta/W system into 3-terminal SOT-MTJ devices, showing a level of performance similar to that of W-based systems. Our results show that orbital physics can be easily integrated into SOT-MTJ systems, offering a viable strategy to enhance SOT-MRAM efficiency. In addition, we propose and demonstrate a proof-of-concept for vertical non-local switching of SOT-MTJ using orbital torques, simplifying bottom-pinned SOT-MRAM fabrication.
\end{abstract}
%TC:endignore

\section{Introduction}
\begin{figure*}[h!]
    \centering
    \includegraphics[width=\linewidth]{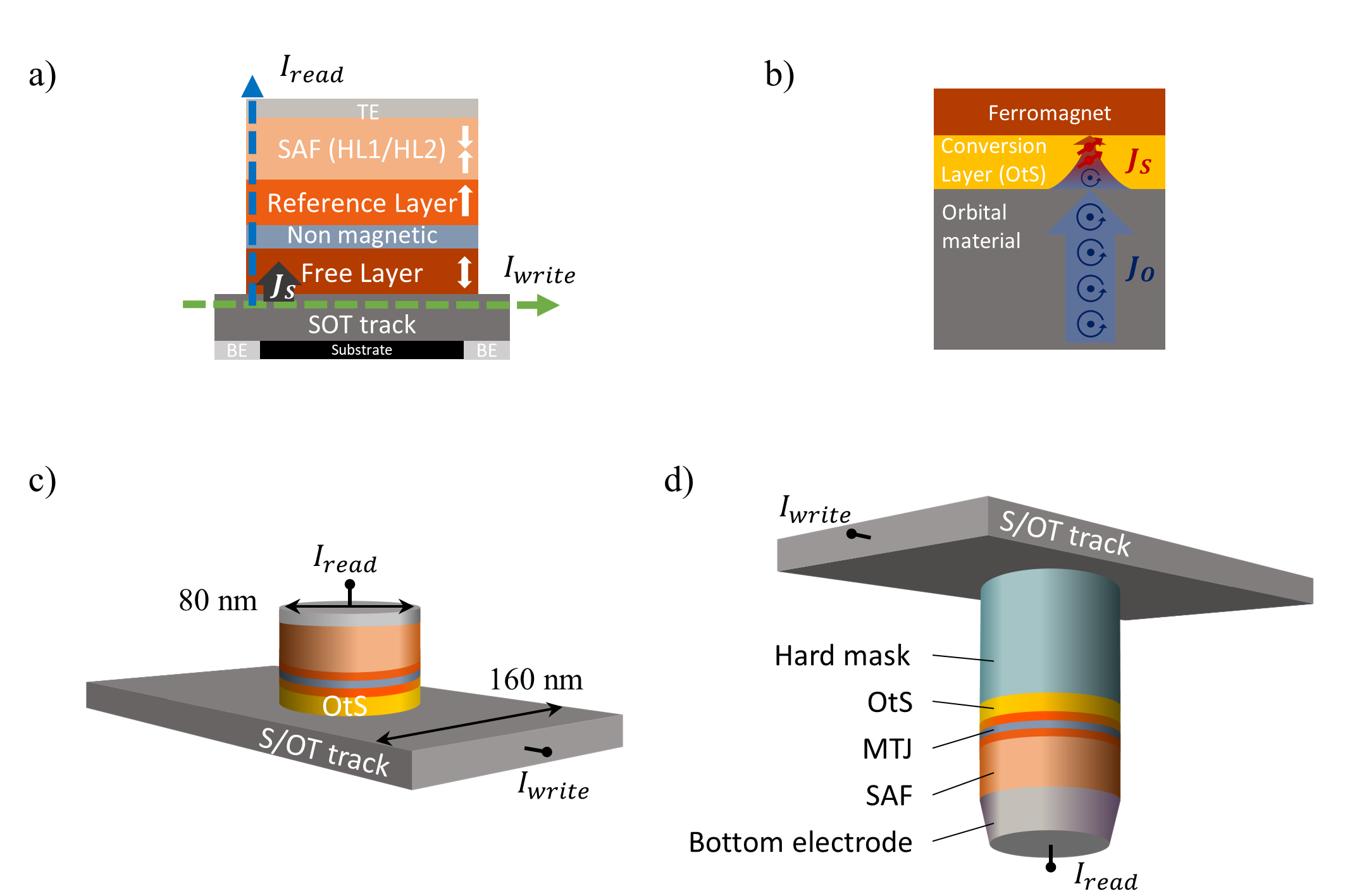}
    \caption{a) Sketch of a 3-terminal SOT-MTJ. The write current is injected in the SOT track, generating a spin current $J_s$ that is transferred to the free layer, while the state of the MTJ is read via TMR. b) Orbital to spin conversion scheme. The orbital current is generated within the orbital material and subsequently converted into a spin current via the orbital-to-spin conversion layer, before propagating into the ferromagnet. c) Schematic view of the devices measured in this work with S/OT track and pillar dimensions. To be noted that in the case of the reference sample the OtS layer is not present. d) Scheme of the proposed bottom-pinned MTJ. The hard mask is used to propagate the orbital and/or spin current into the free layer of the MTJ, simplifying the fabrication process.}
    \label{fig:fig1}
\end{figure*}

The development of electrically controlled nanomagnets for spintronics applications is attracting significant attention, particularly for the realization of non-volatile magnetic memories (MRAMs) \cite{Krizakova2022}. This interest is driven by the pressing challenges faced by the microelectronics industry due to the volatility of CMOS-based cache memory elements, such as SRAM and eDRAM \cite{Molas2021}. Spin-transfer-torque (STT)-MRAM is a low-power, high-speed non-volatile memory technology that has already been commercialized as embedded MRAM (eMRAM) and is currently under development for implementation in last-level cache memories. Spin-orbit torque (SOT)-based magnetic tunnel junctions (MTJs) are considered a next-generation MRAM technology, targeting the replacement of SRAM owing to their fast switching processes and high endurance \cite{Krizakova2022}. In such a structure (Fig. \ref{fig:fig1}a), the spin current controlling the switching is generated by a non-magnetic material, typically a heavy metal (HM), due to the spin Hall effect (SHE) and the Rashba-Edelstein effect (REE) \cite{Dyakonov1971, Sinova2015, Manchon2019}. 

While some important roadblocks have been overcome, such as the realization of deterministic switching without a magnetic field \cite{Garello2018, Wang2018, shao_roadmap_2021}, scaling \cite{VanBeek2023}, and array-level demonstration \cite{Yasin2024}, the technology requirements are still not met: the write current has to be less than 150 $\mathrm{\mu A}$, and the currently available SOT materials show moderate efficiencies, high resistivity, and are limited to relatively low thicknesses. Although giant SOT efficiency materials like topological insulators have already been demonstrated to improve SOT switching efficiency \cite{Pai2018}, their use is still currently hindered by industrial integration process requirements, such as total thermal budget and non-degrading MTJ specifications.

Meanwhile, recent predictions and demonstrations \cite{Go2018, Go2020, Go2024, Salemi2022, Lee2021, Lee2021b, Rothschild2022, Sala2022, Dutta2022} of orbital effects, such as the orbital Hall effect (OHE) and the orbital Rashba-Edelstein effect (OREE), could offer a promising route to improve SOT-MTJ write efficiency. Indeed, they exhibit larger magnitudes and diffusion lengths \cite{Go2018, Choi2023} compared to their spin counterparts, and they are present in a much wider class of materials, including light metals and elements with low resistivity. In an analogous way to SHE and REE, these orbital effects give rise to an orbital current that can propagate into adjacent materials. However, the orbital moment does not directly interact with the spin moment. Hence, to exploit these effects in SOT-MTJ devices, spin-orbit coupling (SOC) is required to convert the orbital current into spin current. This can be achieved by selecting a ferromagnet (FM) with sizable SOC, such as Ni. Another solution, which has the benefit of a less restrictive material choice, is to introduce a strong SOC conversion layer between the orbital source and the ferromagnet (Fig. \ref{fig:fig1}b) \cite{Rothschild2022, Sala2022, Lee2021b, Gupta2025}. In this configuration, the orbital current is converted by the Orbital-to-Spin (OtS) conversion layer and transferred to the ferromagnet. Moreover, if the spacer is a heavy metal, two independent channels of spin current are expected to add up linearly to the total effective spin Hall conductivity, possibly increasing the total effective efficiency of the system $\xi_{DL}=(\xi_{DL}^{HM}+\eta_{LS}^{HM}\theta_{OH}^{OS})$, where $\xi_{DL}^{HM}$ is the damping-like efficiency of the HM, $\eta_{LS}^{HM}$ is the OtS conversion efficiency of the HM, $\theta_{OH}^{OS}$ is the orbital Hall angle of the orbital source material \cite{Lee2021b}. 

Although still debated \cite{Valet2024}, several recent experimental studies have reported additional interactions attributed to orbital current effects, observed using different measurement techniques \cite{Choi2023, Lyalin2023, Kumar2023, Wang2023, Sala2022, Lee2021b, Rothschild2022, Ding2020, Gupta2025}. Experiments have shown how SOC can mediate the transfer of angular momentum to local moments, inducing dynamics of the magnetization and corroborating the role of orbital currents in current-induced torques \cite{Lee2021, Krishnia2023} and their inter-conversion to/from spin currents \cite{Sala2022, Ding2020, Lee2021b, Gupta2025}. 

In this study, we identify a technologically relevant OtS conversion material system, Ta/W, that shows perpendicular magnetic anisotropy (PMA), is back-end-of-line (BEOL) compatible (400 C annealing), and has high efficiency. Ta has the advantage of being compatible with industrial fabrication processes and of exhibiting a large orbital Hall angle $\theta_{OH}$ \cite{Dutta2022, Lee2021}. W has large SOC and a large $\xi_{DL}$ \cite{Pai2012}, resulting in an efficient conversion of the orbital current from Ta. In addition, W and Ta can generate a secondary channel of spin current via SHE, further increasing the total effective current-to-spin conversion efficiency. We systematically quantified the effective damping-like efficiency $\xi_{DL}$ as a function of both the conversion layer thickness and the orbital layer thickness on annealed Hall bar samples. In a second step, we fabricated 3-terminal SOT-MTJ devices with 80 $\mathrm{nm}$ pillar diameter devices, and we benchmarked the switching performance of the Ta\slash W material system (Fig. \ref{fig:fig1}c) against our standard W-based SOT-MTJ. Our results demonstrate not only the successful integration of the OtS conversion scheme in an industrially compatible system but also an enhancement in switching performance. Finally, we extend this work to a proof-of-concept demonstration of vertical non-local switching in SOT-MTJ mediated by orbital torques and discuss its implications for the development of bottom-pinned SOT-MRAM (Fig. \ref{fig:fig1}d).

\section{Results and discussion}
\subsection{Material study}
With the conversion scheme depicted in Fig. \ref{fig:fig1}b, it is not possible to directly disentangle the spin and orbital contributions to the total effective $\xi_{DL}$. Hence, we first systematically quantify, via harmonic Hall voltage method (HHV) \cite{Garello2013}, the $\xi_{DL}$ of the reference stack with only the heavy metal (HM) SOT layer sub\slash\slash W(t)\slash FeCoB(1)\slash MgO(1.25)\slash FeCoB(0.5)\slash Ta(3), and sub\slash\slash Ta(t)\slash FeCoB(1)\slash MgO(1.25)\slash FeCoB(0.5)\slash Ta(3) (numbers are layer thicknesses in nm) as a function of the HM layer thickness $t_{W,Ta}$, and we compare it to the $\xi_{DL}$ quantified in the orbital stacks sub\slash\slash Ta(t)\slash W(1.5)\slash FeCoB(1)\slash MgO(1.25)\slash FeCoB(0.5)\slash Ta(3), and sub\slash\slash Ta(20)\slash W(t)\slash FeCoB(1)\slash MgO(1.25)\slash FeCoB(0.5)\slash Ta(3) as a function of $t_W$ or $t_{Ta}$.
\begin{figure*}[h!]
    \centering
    \includegraphics[width=\linewidth]{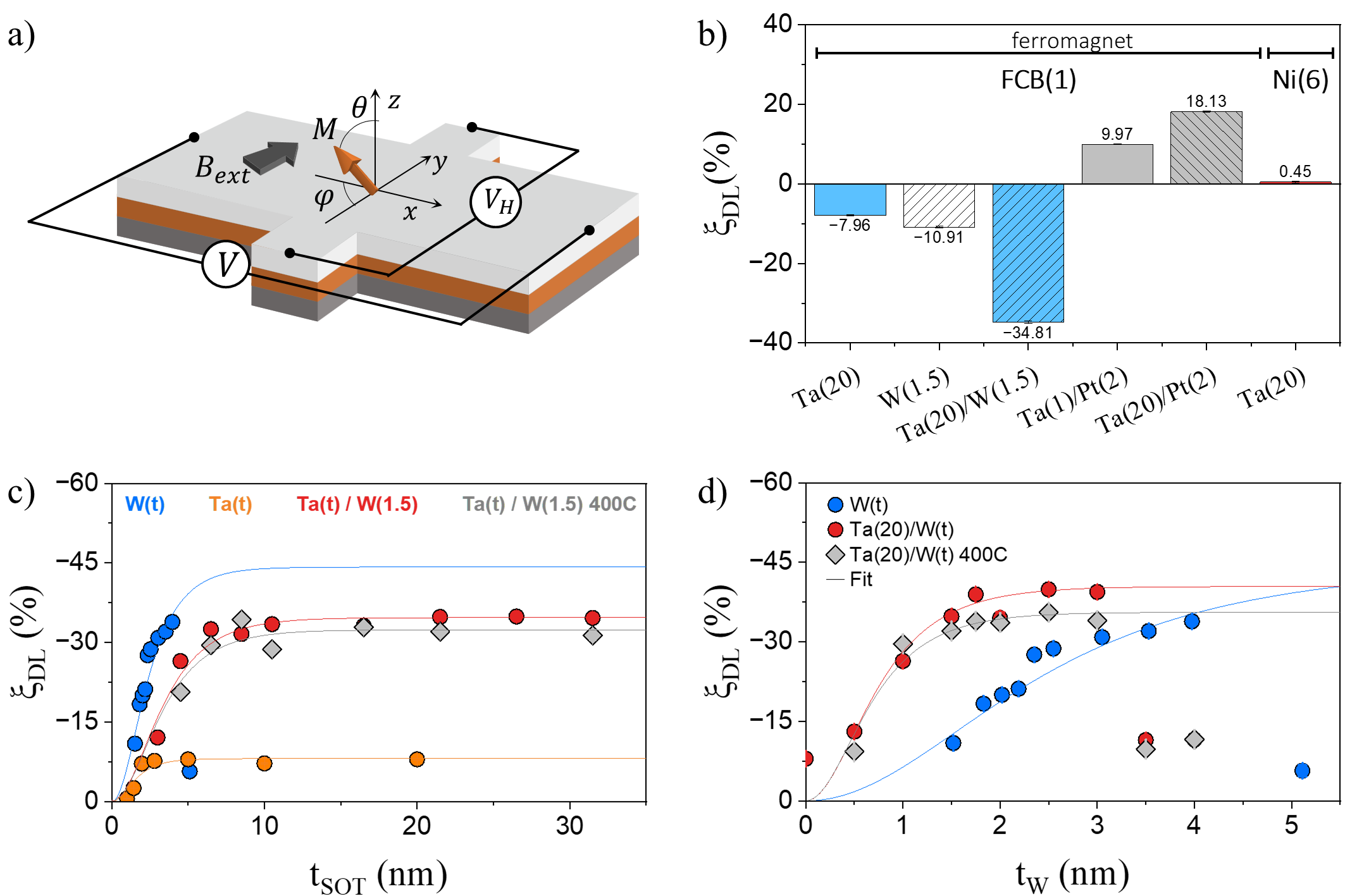}
    \caption{a) Harmonic Hall voltage measurement scheme. An a.c. current $I_{a.c.}$ is injected in the Hall bar, and the Hall voltage is measured as a function of the angle $\varphi$ of the external in-plane field $B_{ext}$. b) Damping-like efficiency ($\xi_{DL}$) for different stack compositions: Ta(20), W(1.5), Ta(20)\slash W(1.5), Pt(2), Ta(20)\slash Pt(2), and Ta(20)\slash Ni(6) (ferromagnet is FeCoB(1) if not specified otherwise). Ta and W show negative spin Hall conductivity while it is positive for Pt. Notably, when the ferromagnet is changed from FeCoB to Ni for the Ta sample, also $\sigma_{SH}$ changes sign, indicating the presence of an additional contribution that can be attributed to orbital physics. c) $\xi_{DL}$ dependence on the S/OT material thickness for W($t_{W}$), Ta($t_{Ta}$), and Ta($t_{Ta}$)\slash W(1.5) after $300^\circ C$ annealing. The Ta/W efficiency is also shown for the annealed samples at $400^\circ C$. d) $\xi_{DL}$ dependence on the SOT material thickness for W($t_{W}$), and Ta(20)\slash W($t_{W}$) after $300^\circ C$ annealing (also $400^\circ C$ for Ta/W). Numbers in parentheses are layer thickness in $\mathrm{nm}$.}
    \label{fig:fig2}
\end{figure*}
All samples were annealed at $300^\circ$C for 10 minutes in a rapid thermal annealing stage under vacuum. The OtS samples (Ta\slash W) were additionally annealed at $400^\circ C$, a requirement to fulfill BEOL process requirements, as well as revealing its impact on $\mathrm{M_S}$, $\mathrm{B_k}$, and $\xi_{DL}$. 

The measurement geometry for the HHV method is illustrated in Fig. \ref{fig:fig2}a). A 10 Hz a.c. current is applied along the Hall bar channel, and the sample can be rotated in a variable external magnetic field $B_{ext}$. All measurements are performed at a constant current density $J_{in}=2$ MA/cm\textsuperscript{2}, calculated by considering the total thickness of the conductive layers (i.e., also considering the FeCoB(1)). From the first and second harmonic resistance, one can quantify the effective damping-like SOT efficiency $\xi_{DL}$ and the effective field-like efficiency $\xi_{FL}$. $\xi_{DL, FL}$ are defined by normalizing $B_{DL, FL}$ by the injected current density $J_{in}$:
\begin{equation}
    \xi_{DL,FL}=\frac{2e}{\hbar}\frac{M_S t_{FM}}{J_{in}}B_{DL,FL}
\end{equation}
Where $B_{DL,FL}$ are the experimentally measured damping-like and field-like effective fields. The $\xi_{FL}$ are reported in the supplementary information. The $\xi_{DL}$ of the reference samples W, Ta, and the Ta/W OtS system is summarized as a function of the SOT track thickness ($t_{SOT}$ = $t_W$, $t_{Ta}$, $t_{Ta+W(1.5)}$) in Fig. \ref{fig:fig2}c. The data are fitted with the drift-diffusion equation (solid lines) $\xi_{DL}(t_{SOT})=\xi_{sat}[1-sech(t_{SOT}/\lambda)]$ \cite{Manchon2019, Liu2011, Sala2022} to estimate the spin diffusion length $\lambda$ and the saturation value $\xi_{sat}$, resulting in $\lambda_W=1.75 \ \mathrm{nm}$, $\lambda_{Ta}=1.14 \ \mathrm{nm}$, $\lambda_{Ta(t)/W}=2.45 \ \mathrm{nm}$, $\xi_{sat}^{W}=-44.2 \%$, $\xi_{sat}^{Ta}=-9.9\%$, and $\xi_{sat}^{Ta(t)/W}=-34.7\%$. It is worth noting that these values are obtained under the assumption of transparent interfaces and therefore underestimate the intrinsic value of $\xi_{DL}$.

Interestingly, $\xi_{DL}$ increases with increasing $t_{SOT}$ in all systems, suggesting that the main current-to-spin conversion mechanism is of bulk origin (i.e., SHE/OHE). However, the Ta(1.5)\slash W(1.5) sample shows considerably lower resistivity compared to the thicker Ta(t)\slash W(1.5) samples ($\rho_{Ta(1.5)/W(1.5))} \sim 84 \ \mathrm{\mu \Omega cm}$ and $\rho_{Ta(t)/W(1.5)} \sim 140\ \mathrm{\mu \Omega cm}$), indicating that the W in our samples stabilizes in a mixed $\alpha$/$\beta$ phase when grown on a thin Ta underlayer. This is further confirmed by resistivity maps performed on Ta\slash W wedges, showing that the W transition happens at lower thicknesses when Ta is thinner (supplementary information). Since the $\xi_{DL}$ thickness dependence is influenced by this phase-transition, this adds uncertainty in identifying the role of bulk\slash interfacial (OHE/OREE) as the origin of the quantified torque.

Nevertheless, we find that the Ta\slash W(1.5) system attains the same values of $\xi_{DL}$ as the W reference system. These results are surprising if we take into account that we are using only 1.5 nm of W as a conversion layer, which shows a $\xi_{DL}=10.9\%$. Consequently, if we compare the DL efficiency of the W(1.5), Ta(20), and Ta(20)\slash W(1.5) samples, the efficiency in the Ta\slash W system is enhanced by a factor of 3.2 with respect to the W(1.5) sample and by 4.4 with respect to the Ta(20) sample. Importantly, these samples were also annealed at $400^\circ\text{C}$, and the efficiency values of the orbital stack were found to be only marginally affected, with $\xi_{sat, Ta(t)/W}^{400^\circ C}=-32.4 \%$ and $\lambda_{Ta(t)/W}^{400^\circ C}=2.41 \ \mathrm{nm}$. Meanwhile, the anisotropy is significantly enhanced, from $B_k \approx 0.5 \ \mathrm{T}$ to $B_k \approx 1.1 \ \mathrm{T}$, which translates to improved thermal retention and PMA at the device level.

The sign of $\xi_{DL}$ is found to be negative, although $\theta_{OH}$ is predicted to be positive for Ta and W. This points to the fact that the orbital current is converted with the sign of the SOC of the conversion material, which is negative for W \cite{Salemi2022, Go2024, Ding2020, Sala2022, Lee2021}. This is confirmed when substituting W with Pt as the conversion layer; we observe an opposite sign of $\xi_{DL}$ corresponding to the sign of SOC in Pt, and we consistently measure an enhancement of $\xi_{DL}$ in Ta(20)\slash Pt(2) compared to the Pt reference sample (Fig. \ref{fig:fig2}b).

To further identify the presence of orbital effects in Ta, we fabricated a Ta(20)\slash Ni(6) sample. Ni is known for its relatively large positive SOC, which should allow for the conversion of orbital current without requiring an OtS layer \cite{Lee2021b, Lee2021, Go2020, Amin2019, Dutta2022}.  In this scenario, the measured efficiency should reflect the interplay between the negative spin Hall effect (SHE) in Ta and the positive OtS conversion in Ni. The fact that our measurements show a small but positive  $\xi_{DL}$ suggests that, within the two-channel current model picture, orbital effects in Ta are significant and that the OtS conversion by Ni of orbital current in Ta dominates over the SHE generated in Ta.

Then, we investigated the dependence of $\xi_{DL}$ on W thickness in Ta(20)/W(t) samples, as shown in Fig. \ref{fig:fig2}d. The efficiency is consistently enhanced with respect to the reference samples. In particular, we observe a maximum value of $\xi_{DL}$ ($\xi_{DL} = -39.90\%$ for $t_W = 2.5 \ \mathrm{nm}$) in the Ta\slash W system, which is 18\% larger than the highest value of $\xi_{DL}$ in the reference samples ($\xi_{DL} = -33.87\%$ for $t_W = 4 \ \mathrm{nm}$). This can be interpreted as the constructive combination of the SHE generated by W and OHE\slash OREE generated by Ta. Meanwhile, the shallow decrease of $\xi_{DL}^{Ta/W}$ at about $t_W$ = 1.75 nm can be understood by considering that when the thickness of the conversion layer exceeds the spin diffusion length in W, scattering events with spin-flips become dominant, and the spin-orbital conductivity saturates to a finite value determined by the SHE in the conversion layer \cite{Sala2022}. However, it is difficult to draw clear conclusions as there is a sharp drop at $t_W$ = 3 $\mathrm{nm}$, which is due to the $\beta$-to-$\alpha$ transition phase of W, known to occur around 4 nm for annealed W on SiO\textsubscript{x}, and to lower thicknesses when W is grown on a thin metallic seed \cite{McHugh2020, Vudya2021}. In the case of $400^\circ C$ annealing, we report similar trends to the $300^\circ C$ case, but with a slight reduction in efficiency values of about 13\% in $1 \ \mathrm{nm} < t_W < 3 \ \mathrm{nm}$, and a more pronounced plateau. 

Finally, we tentatively separate the contributions from the SHE and OHE in the orbital stacks by means of a simple current distribution model in the saturation region for the Ta(20)\slash W(1.5) sample. We first estimate the resistivity of W and Ta by linearly fitting the conductance over the variation of the layer thickness in W(t), Ta(t), Ta(t)\slash W, and Ta\slash Pt(t) sample series, $G_{tot}=\frac{\sigma_X w}{l} t_X + G_{rest}$,  where X = W, Ta, Pt, $G_{rest}$ is the conductance of the fixed thickness layers in the samples, $w$ is the Hall bar width, and $l$ is the distance between the Hall bar arms used to monitor sample resistivity. Using the quantified efficiencies of W(1.5)\slash FeCoB and Ta(20)\slash FeCoB in the reference samples, one can simply estimate the effective SOT field $B_{DL}=\theta_{DL}\hbar J_c^{layer}/(2eM_St_{FM})$ generated by each layer in Ta(20)\slash W(1.5)\slash FeCoB(1) by SHE. By subtracting the sum of the two from the experimentally measured SOT field, we estimated the additional contribution supposedly generated by the OtS conversion in Ta(20)\slash W(1.5), as reported in Table \ref{tab:tab1}.
\renewcommand{\arraystretch}{2}
\begin{table*}[]
    \centering
    \resizebox{\textwidth}{!}{%
    \begin{tabular}{c c c c c c c c c}
         Stack & Layer & $\rho_{R_{//}} \ [\mathrm{\mu\Omega cm}]$ & $I_{stack}\ [\mathrm{mA}]$ & $J^{layer}_{R_{//}}$ & $\theta_{DL}^{intr}$ [\%]& $B_{DL}^{est}$ [mT] & $B_{DL}^{exp}$ [mT] & $B_{DL}^{OT}$ [mT] \\ \hline \hline
         W(1.5)/FeCoB(1) & W(1.5) & $138.9 \pm 6.3$& 0.50 & 1.51& -14.45& & & \\
         Ta(20)/FeCoB(1) & Ta(20) & $145.7 \pm 2.6$& 4.20 & 1.94& -8.20& & & \\
         Ta(1)/Pt(2)/FeCoB(1) & Pt(2) & $67.5 \pm 17.3$& 0.80 & 2.93& 6.93 & & & \\ \hline
         \multirow{2}{*}{Ta(20)/W(1.5)/FeCoB(1)}& W(1.5) & & \multirow{2}{*}{4.50} & 2.01& & -0.87& \multirow{2}{*}{-2.30} & \multirow{2}{*}{$-1.20 \pm 0.24$}\\
         & Ta(20) & & & 1.92 & & -0.47& & \\ \hline
         \multirow{2}{*}{Ta(20)/Pt(2)/FeCoB(1)}& Pt(2) & & \multirow{2}{*}{4.60} & 3.87 & & 0.57 & \multirow{2}{*}{0.77} & \multirow{2}{*}{$0.36 \pm 0.16$} \\
         & Ta(20) & & & 1.79 & & -0.31 & & \\
         \hline\hline
    \end{tabular}
    }
    \caption{Estimated $\rho_{R_{//}}$ from conductance fit, estimated intrinsic spin Hall angle $\theta_{DL}^{intr}$ for reference samples, and $B
    _{DL}$ contribution separation for OtS samples by means of a simple parallel resistor model.}
    \label{tab:tab1}
\end{table*}
\renewcommand{\arraystretch}{1}

In the Ta(20)\slash W(1.5) sample, we estimate the damping-like effective fields as $B_{DL}^{W} = -0.87\ \mathrm{mT}$ and $B_{DL}^{Ta} = -0.47\ \mathrm{mT}$. Even assuming complete transmission of the Ta-generated spin current through the W layer to the ferromagnet, the combined Ta and W contributions are insufficient to account for the measured SOT effective field $B_{DL}^{exp} = -2.30\ \mathrm{mT}$. This discrepancy implies an additional contribution of $B_{DL}^{OT} = -1.20 \pm 0.24\ \mathrm{mT}$, where the uncertainty boundaries correspond to the cases in which the spin current from Ta is either fully transmitted to the FeCoB or completely absorbed by the W layer. Therefore, such a large enhancement cannot be attributed solely to the SHE; rather, it could be explained by orbital current generation in Ta and its subsequent conversion into spin current in W.

Finally, we estimated the individual layers SHE contribution in Ta(20)\slash Pt(2)\slash FeCoB(1). Here, we used a series of Ta(1)\slash Pt(t)\slash FeCoB(1) as reference samples, where the layer Ta(1) serves as a seed for better growth of Pt; noting that Ta(1) showed negligible SOT efficiency. In this case, we also find an additional positive contribution of $B_{DL}^{OT}=0.36 \pm 0.16 \ \mathrm{mT}$, consistent with the OtS conversion scenario.

It is important to note that this model neglects interfacial spin transparency as well as interfacial phenomena such as spin memory loss and spin backflow \cite{Berger2018, Dolui2017, Han2021}. When W is grown on Ta, the resulting W(1.5)\slash FeCoB interface may differ from that obtained when W is deposited on SiO$_2$, potentially leading to enhanced spin transmission in the Ta/W system due to better interfacial quality. However, we observe that: (i) the enhancement in damping-like efficiency is observed in both the Ta/W system and the Ta/Pt system; (ii) the sign of $\xi_{DL}$ is consistent with the sign of the spin Hall angle of the conversion layer; (iii) in the Ta/Ni system, a positive DL efficiency is observed, in agreement with the sign of the SOC of Ni. These results point more strongly to an orbital-to-spin conversion scenario.

For the field-like torque (supplementary information), surprisingly, we do not observe a reduction as expected when W turns into alpha, suggesting the presence of a non-negligible spin accumulation at the W/FCB interface that is independent of W.

\subsection{Device switching characterization}
\begin{figure*}[h!]
    \centering
    \includegraphics[width=\linewidth]{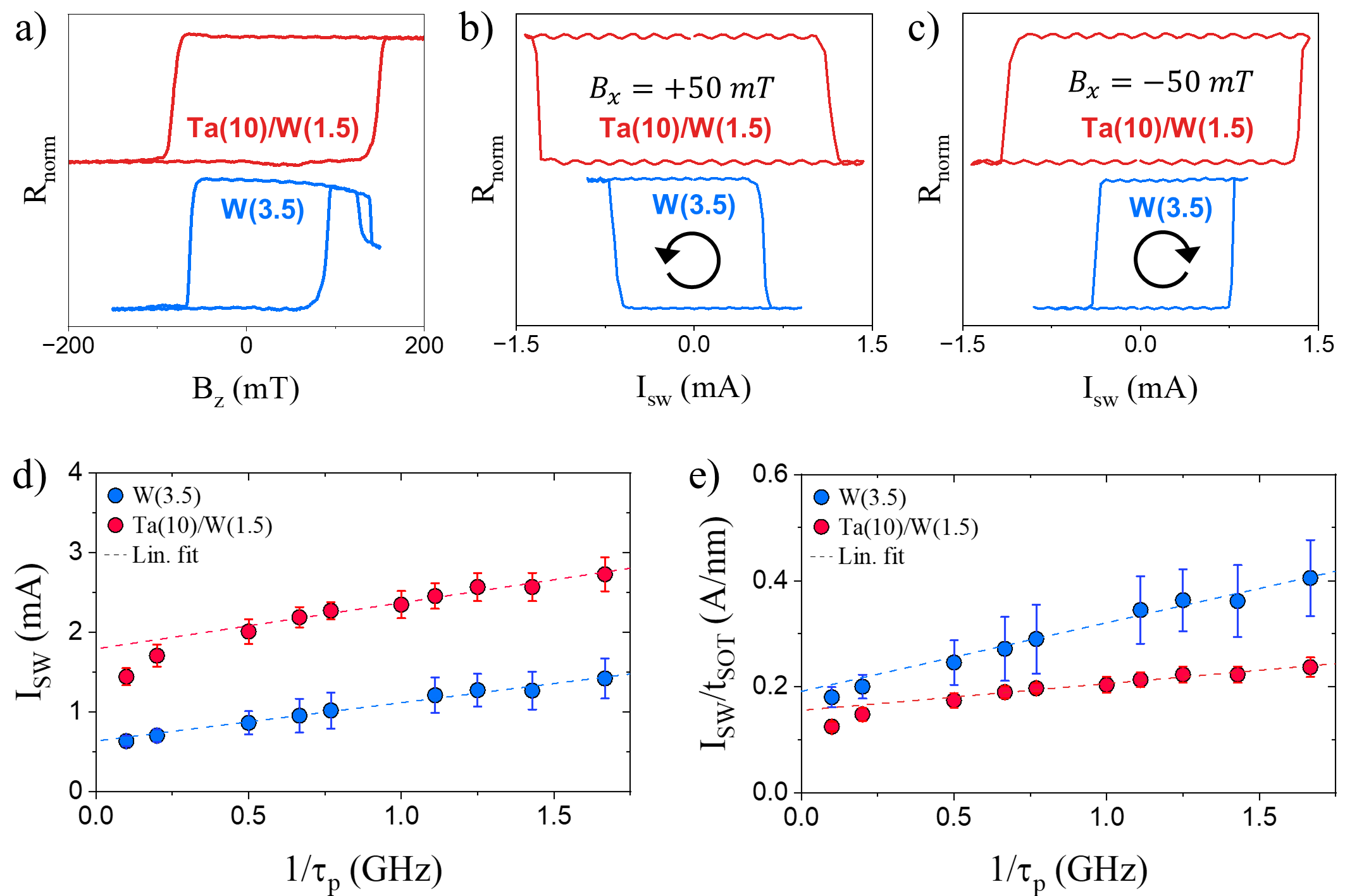}
    \caption{a) 3-terminal SOT-MTJ devices R-H loop, for reference sample W(3.5)\slash FeCoB(1) and OtS sample Ta(10)\slash W(1.5). b) Device switching plot for  $\tau_P=10 \ \mathrm{ns}$ and $B_x = \pm 50 \ \mathrm{mT}$, for reference sample W(3.5)\slash FeCoB(1) and OtS sample Ta(10)\slash W(1.5)\slash FeCoB(1). Also OtS sample shows SOT switching symmetry. c) Switching current $I_{SW}$ dependence on the inverse of the applied pulse length $\tau_p$, for reference and OtS samples. The data are fitted in the intrinsic regime by means of a linear model (see text). d) Normalized $I_{SW}$ on the SOT track thickness below the MTJ $t_{SOT}$ as a function of the inverse of the applied pulse length. Notably, the Ta(10)\slash W(1.5) samples shows lower switching current values.}
    \label{fig:fig3}
\end{figure*}

Building on blanket sample findings, we fabricated three-terminal MTJs that include W as a reference material and an OtS stack to assess its switching performance at the device level (Fig.\ref{fig:fig1}c). The OtS sample consists of Sub\slash \slash Ta(10)\slash W(1.5)\slash FeCoB(1)\slash MgO(1.25)\slash FeCoB(1)\slash SAF, while the reference sample composition is sub\slash \slash W(3.5)\slash FeCoB(1)\slash MgO(1.25)\slash FeCoB(1)\slash SAF. In the reference sample, we tuned the etching conditions to stop the pillar etch at the SOT line. In the OtS sample, one of the main advantages is the larger tolerance to the etching margins; therefore, we tuned the etching conditions to leave Ta(7) as the SOT track. The TMR is found to be between 10\% and 20\% for both W and Ta/W samples. Although weak, it relates more to our MgO stack quality and our fabrication process than to the limitations of these material systems.

Fig. \ref{fig:fig3}a shows that OtS samples have a significantly improved free layer coercive field $B_c = 116.4 \ \mathrm{mT}$, almost twice as large as that of the W samples. Then, in Fig. \ref{fig:fig3}b,c, we present exemplary switching loops for $\tau_P=10 \ \mathrm{ns}$ and $B_x = \pm 50 \ \mathrm{mT}$. Both samples switch with a clear SOT character, i.e., the polarity depends on the combination of the current and the applied in-plane field polarities. Finally, we characterized the switching current as a function of the applied pulse length down to 500 ps, and we report in Fig. \ref{fig:fig3}d the median switching current $I_{SW}$ averaged over five devices (obtained from 100 loops per device) as a function of the inverse of the pulse length $1/\tau_p$. One can estimate from this measurement the critical current in the intrinsic switching regime, following the model $I_c = I_{c0} + q/\tau_p$, where $I_{c0}$ is the intrinsic critical switching current, and q is an effective charge parameter that determines the rate at which the angular momentum is transferred to the free layer (FL) \cite{Garello2014}. By fitting the data for $\tau_p<1 \mathrm{ns}$, we find $I_{c0}^{W(3.5)}=0.67 \ \mathrm{mA}$ and $I_{c0}^{Ta(10)/W(3.5)}=1.86 \ \mathrm{mA}$, respectively. However, it should be noted that the Ta/W sample exhibits substantially higher coercivity ($B_C^{W(3.5)}=63.17 \ \mathrm{mT}$, and $B_C^{Ta(10)/W(1.5)}=116.4 \ \mathrm{mT}$) and anisotropy ($B_k^{W(3.5)}=163.8 \ \mathrm{mT}$, and $B_k^{Ta(10)/W(1.5)}=232.0 \ \mathrm{mT}$), which implies that a lower switching current density is intrinsically required for the W samples. However, the SOT track below the MTJ is nearly four times thicker for the Ta/W sample, which inherently leads to higher absolute switching currents. Hence, to compare the current-to-spin conversion efficiency of the two systems, we normalize $I_{SW}$ by the coercivity of the samples and the SOT track thickness below the MTJ (Fig. \ref{fig:fig3}e). In this case, both systems exhibit comparable values for $I_{c0}/B_{C}$, of $I_{c0}^{W(3.5)}/B_C^{W(3.5)}=0.011 \ \mathrm{mA/mT}$ and $I_{c0}^{Ta(10)/W(1.5)}/B_C^{Ta(10)/W(1.5)}=0.016 \ \mathrm{mA/mT}$. Furthermore, after normalization by the thickness of the SOT track below the pillar, the Ta/W sample achieves a lower $I_{c0}$ than W, particularly $I_{c0}^W/t_W=0.19 \ \mathrm{mA/nm}$ and $I_{c0}^{Ta/W}/t_{Ta/W}=0.16 \ \mathrm{mA/nm}$. While the two systems show similar switching performances, the Ta/W system provides superior scalability and improved process compatibility in terms of thermal budget (annealing) and tolerance to pillar etching, easing integration and development on a large scale and industrial platforms.

\subsection{Towards bottom-pinned SOT-MTJ}
\begin{figure*}[h!]
    \centering
    \includegraphics[width=\linewidth]{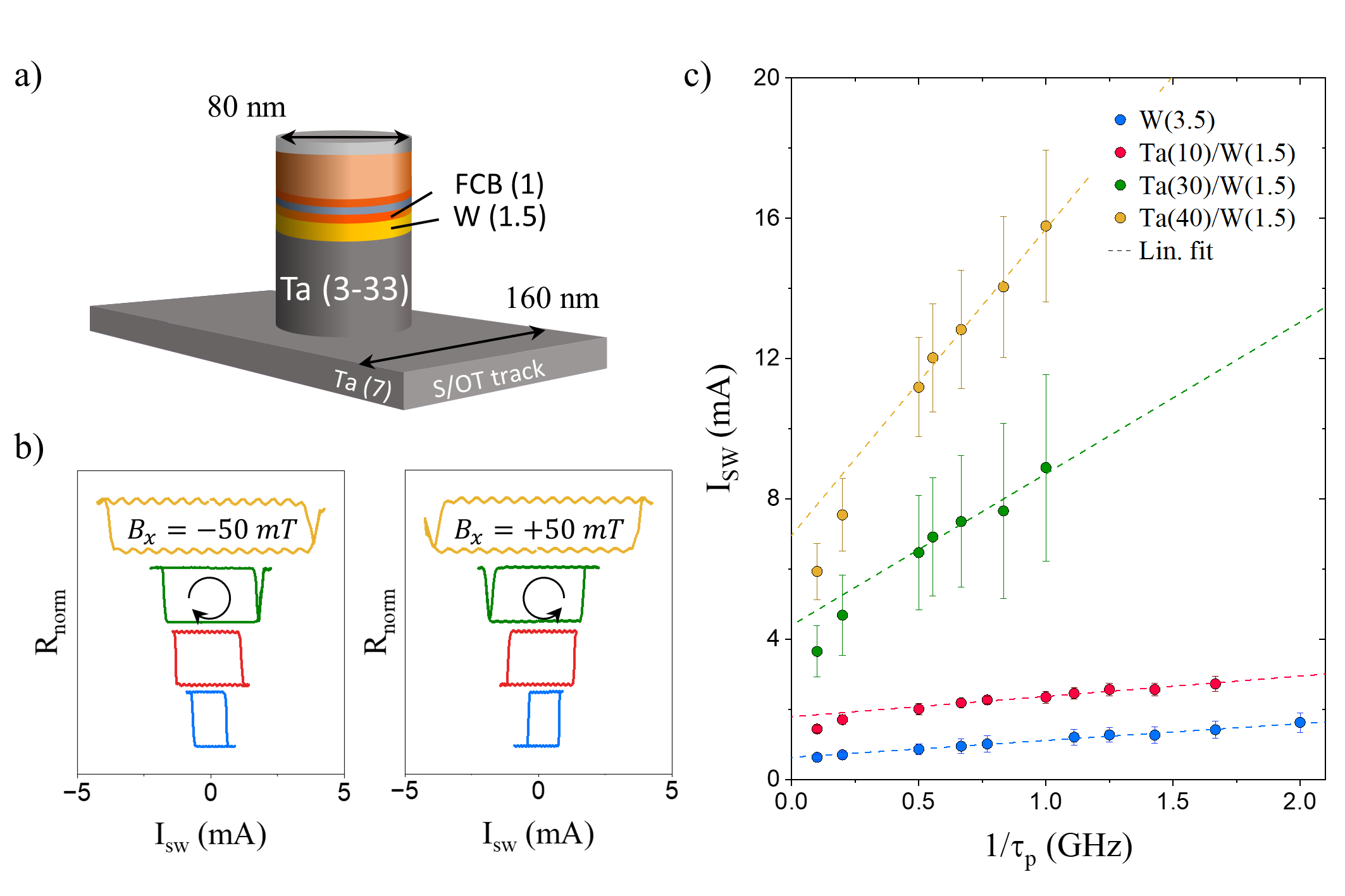}
    \caption{a) Device scheme. The etch is stopped inside the SOT track such as to leave a thick Ta spacer layer between the S/OT track where the current is injected and the OtS/free layer. b) Device switching plots for  $\tau_P=10 \ \mathrm{ns}$ and $B_X = \pm 50 \ \mathrm{mT}$, for samples W(3.5), Ta(10)\slash W(1.5), Ta(30)\slash W(1.5), and Ta(40)\slash W(1.5). All samples switch with SOT symmetry. c) $I_{SW}$ dependence on the inverse of the applied pulse length $\tau_p$. The data are fitted in the intrinsic regime by means of a linear model (see text). Notably, switching is observed for all samples. As expected, Ta(30)\slash W(1.5), and Ta(40)\slash W(1.5) samples require higher current to switch due to the higher thickness. This is attributed to both the decay of orbital currents in the Ta spacer, and the reduced current in the W layer.}
    \label{fig:fig4}
\end{figure*}

Finally, we leverage our findings to introduce a novel concept of vertical non-local switching in SOT-MTJ devices, which enables a new approach that could greatly simplify the fabrication process of ``bottom-pinned'' SOT-MTJs. The bottom-pinned design is classically used in commercialized STT-MRAM technology. It consists of designing the stack by placing the FL on top and the SAF at the bottom, which is known to improve MTJ properties by reducing stress on the FL/MgO while additionally improving the yield of the device \cite{Krizakova2022, Liu2017}. Although the bottom-pinned SOT-MTJ was recently demonstrated \cite{Li2023, Huang2025}, its fabrication requires introducing several critical steps, such as precisely removing the MTJ hard mask without affecting FL properties and depositing sufficiently conformal SOT material. This would hinder its implementation into industrial processes and limit its scaling.

We propose an alternative approach inspired by lateral spin valve (LSV) devices \cite{Gao2025, Valenzuela2006}, which is a spin current metrology method in which the spin current injector and detector are spaced hundreds of nanometers apart. It was also demonstrated that it is possible to exploit the non-local (NL) spin current drift to switch magnetic nano-pillars  \cite{yang2008} and 3-terminal STT devices \cite{Sun2009}. We transposed this in-plane NL switching concept to vertical 3-terminal SOT-MTJ devices to introduce a simplified integration scheme (Fig. \ref{fig:fig1}d) of the ``bottom-pinned" MTJ \cite{NLpatent}. In analogy with LSV, the spacer layer is the metallic hard mask used to etch the MTJ pillar, the detector is the MTJ free layer, and the spin/orbital current generator is the S/OT line. This design simplifies the fabrication process, introducing minimal changes to the current STT process flow while preserving the benefits of both the bottom-pinned MTJ and the 3-terminal MTJ configuration. Moreover, it has the advantage of being less restrictive on material choice and offers the possibility to exploit novel physics, such as orbital current.
\begin{figure*}[h]
    \centering
    \includegraphics[width=\linewidth]{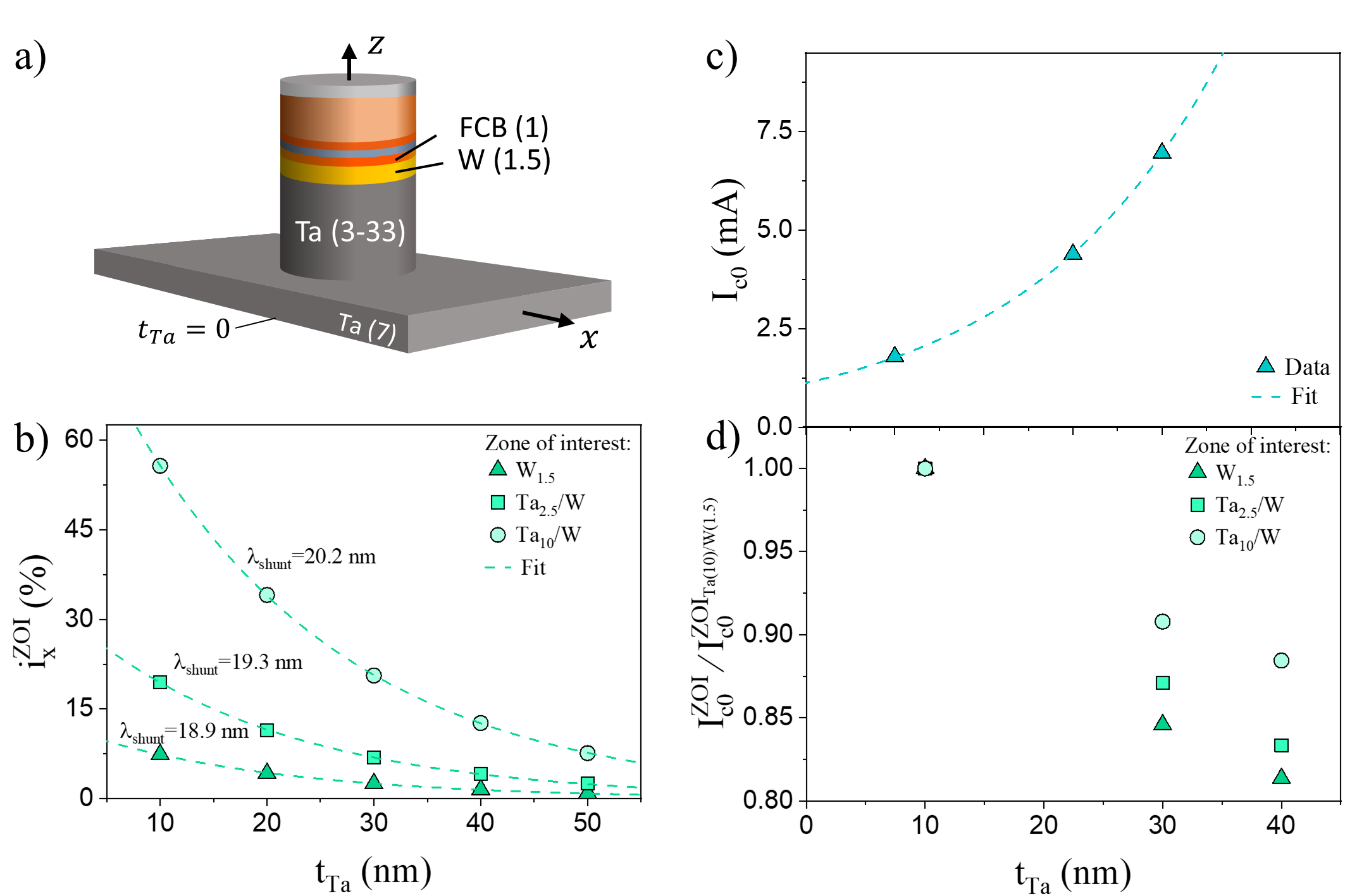}
    \caption{a) Illustration of the simulated and measured MTJ devices. b) Simulated current $x$-component percentage flowing in W(1.5), Ta(2.5)/W(1.5) and Ta(10)/W(1.5) zone of interest (ZOI) for a given current input, as a function of the total Ta thickness $t_{Ta}$. c) Experimentally measured intrinsic switching current $I_{c0}$ as a function of $t_{Ta}$. d) Normalized current fraction at the experimental critical switching current flowing in the zones of interest as a function of $t_{Ta}$. The critical current fraction $I_{c0}^{\mathrm{ZOI}}= i_{x}^{\mathrm{ZOI}} \times I_{c0}$ is normalized by $I_{c0}^{\mathrm{ZOI}_{Ta(10)/W(1.5)}}= i_{x}^{\mathrm{ZOI}_{Ta(10)/W(1.5)}} \times I_{c0}^{Ta(10)/W(1.5)}$ that is the fraction of $I_{c0}$ flowing in the ZOI in the Ta(10)\slash W(1.5) sample. The dotted lines in b) and c) are the exponential fit from which we extract the characteristic lengthscale of the shunting effect and the lengthscale over which $I_{c0}$ increases in the measured devices.}
    \label{fig:fig5}
\end{figure*}
Given the limitations of our cleanroom for integrating 3-terminal bottom pinned MTJ, we fabricated additional Ta(30)\slash W(1.5) and Ta(40)\slash W(1.5) MTJ samples in a top-pinned configuration to explore the feasibility of the non-local control of S/OT-MTJ (Fig. \ref{fig:fig4}a). 80 $\mathrm{nm}$ diameter MTJ pillars were etched down to Ta = 7 $\mathrm{nm}$, separating the S/OT layer and the free layer by 23 (33) $\mathrm{nm}$ Ta spacers. Fig. \ref{fig:fig4}b clearly shows SOT symmetry switching in both sample sets.

We report in figure \ref{fig:fig4}c the median switching current $I_{SW}$ averaged over five devices (obtained from 100 loops per device) as a function of the inverse of the pulse length $1/\tau_p$. Notably, we observe switching of the free layer down to 1 $\mathrm{ns}$ pulse duration, after which we are limited by the maximum applicable voltage in our experimental setup. The critical switching current $I_{c0}$ increases with the thickness of the patterned Ta pillar. This is expected to be due to the reduced current flowing in the W and at the Ta\slash W interface, which decreases the contributions from the SHE of W and the OREE to the total spin current, respectively.

To determine whether the increase in $I_{c0}$ can be explained solely by the shunting effect and associated reduced current in W, or whether an additional long-range bulk contribution (possibly arising from the OHE) is present, we performed COMSOL simulations to estimate the shunting length in the Ta\slash W system. Here, the shunting length $\lambda_{shunt}$ is the characteristic electrical length-scale describing the current redistribution and partial leakage in the MTJ pillar due to the shunting effect. We quantified from simulations $i_{x}^{\mathrm{ZOI}}$, the current percentage flowing in the zones of interest (ZOI). The ZOI includes the possible ``effective volumes'' contributing to switching; i.e., the W(1.5) layer alone, the Ta(2.5)/W(1.5), and Ta(10)/W(1.5). To differentiate them from the sample nomenclature, we denote them as W$_{1.5}$, Ta$_{2.5}$\slash W, and Ta$_{10}$\slash W. In the simulations, we used the resistivity values estimated from the conductance fit reported in Tab. \ref{tab:tab1}. The results of the simulations are reported in Fig. \ref{fig:fig5}b. By fitting the decrease of $i_{x}^{\mathrm{ZOI}}$ with a current decay diffusion model $i_{x}^{\mathrm{ZOI}}(t_{Ta}) = Ae^{-t_{Ta}/\lambda_{shunt}}$ \cite{Berger1972, Gao2025}, we extract a shunting characteristic length-scale of $\lambda_{shunt}^{W_{1.5}} = 18.9 \ \mathrm{nm}$, $\lambda_{shunt}^{Ta_{2.5}/W} = 19.3 \ \mathrm{nm}$, and $\lambda_{shunt}^{Ta_{10}/W} = 20.2 \ \mathrm{nm}$.

We then fit the experimental values of $I_{c0}$ with $I_{c0}(t_{Ta}) = B e^{t_{\mathrm{Ta}}/\lambda_{\mathrm{exp}}}$ (Fig.\ref{fig:fig5}c), where $\lambda_{\mathrm{exp}}$ represents an effective characteristic length-scale of the system to be compared with the simulated value $\lambda_{\mathrm{shunt}}$. The fit yields $\lambda_{\mathrm{exp}} = 22.0\,\mathrm{nm}$. Although only three data points are available, the exponential dependence provides a reasonable description of the data, indicating that the shunting effect is the dominant factor responsible for the increase in $I_{c0}$. It is worth noting that the simulation with COMSOL neglects the effects of the interfaces and other non-idealities in our system; therefore, it overestimates the real shunting length. We find $\lambda_{exp} > \lambda_{shunt}$ of 16.4\% (14\%) when considering W$_{1.5}$ (Ta$_{2.5}$\slash W) as an effective switching volume. Interestingly, we still find $\lambda_{exp} > \lambda_{shunt}$ of about 9\% even if we consider 10 nm of Ta to contribute to the switching.

To further confirm the presence of an additional long-range contribution, we calculated the reduction of the critical switching current in the ZOI in the different Ta thickness samples, calculated as $I^{\mathrm{ZOI}}_{c0}=i_{x}^{\mathrm{ZOI}} \times I_{c0}$. Fig \ref{fig:fig5}d shows $I^{\mathrm{ZOI}}_{c0}$ normalized by $I_{c0}^{\mathrm{ZOI}_{Ta(10)/W(1.5)}}$ (i.e., the current flowing in the ZOI in the Ta(10)\slash W(1.5) sample) as a function of the total Ta thickness below the ferromagnet. We find that the critical switching current for the three ZOI decreases as the Ta thickness increases. In particular, the critical switching current in the W$_{1.5}$ (Ta$_{2.5}$\slash W) layer is about 19\% (17\%) less in Ta(40)\slash W(1.5) than in Ta(10)\slash W(1.5) samples. If switching were governed exclusively by the local current flowing into the ZOI, the critical switching current would be expected to be independent of the Ta thickness. Taken together with the previous observations, these estimates are consistent with the presence of an additional long-range contribution that grows with $t_{Ta}$.

\subsection{Conclusions}
In conclusion, we systematically investigated the thickness dependence of the SOT efficiency for W, Ta, and Ta/W as a function of both W and Ta thicknesses. We found a $\sim 20\%$ enhancement in efficiency for the Ta/W bilayer compared to the reference W samples, which we attribute to orbital physics. We estimated that this additional contribution is comparable in magnitude to the spin Hall effect from W(1.5) and that it represents $\sim50$\% of the total measured $B_{SOT}$. In addition, the Ta/W-based system exhibits large perpendicular magnetic anisotropy and is compatible  with $400^\circ C$ annealing, an important requirement for the MRAM industrial integration process. We successfully transferred this orbital-to-spin conversion scheme into 3-terminal S/OT-MTJ and benchmarked their switching performances against standard W-based devices. While Ta/W systems require more current to switch due to a fourfold increase in thickness, they demonstrate superior switching efficiency compared to W samples when normalized by the thickness below the MTJ, highlighting significant potential for optimization. Finally, we provided indications of long-range orbital diffusion in the MTJ, further supported by the COMSOL current-density distribution simulations. However, the relatively limited orbital diffusion length of the investigated materials remains a key constraint. This suggests a promising pathway toward non-local control of MTJs through orbital transport mechanisms by exploring materials with longer orbital diffusion lengths. 

In conclusion, our results underscore that orbital physics can be harnessed to improve SOT writing performance in SOT–MTJ, and that the Ta\slash W system has the potential to outperform the current standard W-based systems in terms of critical switching current while maintaining fast operation and showing superior MTJ properties, as well as room for SOT track optimization. This achievement opens a promising pathway for a simplified integration of bottom-pinned SOT-MTJs.

\nolinenumbers
%TC:ignore
\begin{acknowledgement}
This work was supported by the France 2030 government investment plan, managed by the French National Research Agency under grant reference PEPR Electronique (ANR-22-PEEL-0009); by the French RENATECH network, implemented on the Upstream Technological Platform in Grenoble (PTA) (ANR-22-PEEL-0015); by the Carnot project PRIME SPOT (ANR P-22-03813); by the Horizon Europe grant 101183277 in the framework of the Chips JU NanoIC Pilot Line; and by ORCHID program (project 52694UJ), funded by the French Ministry for Europe and Foreign Affairs, and for Higher Education and Research and the National Science and Technology Council.
The authors thank Jerome Faure-Vincent and Stephane Auffret for sample deposition and material development, Alain Marty and Gwenael Atcheson for technical support, and YoshiChika Otani for valuable scientific discussions. 

\end{acknowledgement}
%TC:endignore


\bibliography{bibliography}

@ARTICLE{Krizakova2022,
  doi = {10.1016/j.jmmm.2022.169692},
  url = {https://doi.org/10.1016/j.jmmm.2022.169692},
  year = {2022},
  month = nov,
  publisher = {Elsevier {BV}},
  volume = {562},
  pages = {169692},
  author = {Viola Krizakova and Manu Perumkunnil and S{\'{e}}bastien Couet and Pietro Gambardella and Kevin Garello},
  title = {Spin-orbit torque switching of magnetic tunnel junctions for memory applications},
  journal = {Journal of Magnetism and Magnetic Materials}
}

@inproceedings{Garello2018,
  doi = {10.1109/vlsic.2018.8502269},
  url = {https://doi.org/10.1109/vlsic.2018.8502269},
  year = {2018},
  month = jun,
  publisher = {{IEEE}},
  author = {K. Garello and F. Yasin and S. Couet and L. Souriau and J. Swerts and S. Rao and S. Van Beek and W. Kim and E. Liu and S. Kundu and D. Tsvetanova and K. Croes and N. Jossart and E. Grimaldi and M. Baumgartner and D. Crotti and A. Fumemont and P. Gambardella and G.S. Kar},
  title = {{SOT}-{MRAM} 300MM Integration for Low Power and Ultrafast Embedded Memories},
  booktitle = {2018 {IEEE} Symposium on {VLSI} Circuits}
}

@article{Pai2018,
  doi = {10.1038/s41563-018-0146-x},
  url = {https://doi.org/10.1038/s41563-018-0146-x},
  year = {2018},
  month = jul,
  publisher = {Springer Science and Business Media {LLC}},
  volume = {17},
  number = {9},
  pages = {755--757},
  author = {Chi-Feng Pai},
  title = {Switching by topological insulators},
  journal = {Nature Materials}
}

@article{Wang2018,
  doi = {10.1038/s41928-018-0160-7},
  url = {https://doi.org/10.1038/s41928-018-0160-7},
  year = {2018},
  month = nov,
  publisher = {Springer Science and Business Media {LLC}},
  volume = {1},
  number = {11},
  pages = {582--588},
  author = {Mengxing Wang and Wenlong Cai and Daoqian Zhu and Zhaohao Wang and Jimmy Kan and Zhengyang Zhao and Kaihua Cao and Zilu Wang and Youguang Zhang and Tianrui Zhang and Chando Park and Jian-Ping Wang and Albert Fert and Weisheng Zhao},
  title = {Field-free switching of a perpendicular magnetic tunnel junction through the interplay of spin{\textendash}orbit and spin-transfer torques},
  journal = {Nature Electronics}
}

@article{Go2018,
  title = {Intrinsic Spin and Orbital Hall Effects from Orbital Texture},
  volume = {121},
  ISSN = {1079-7114},
  url = {http://dx.doi.org/10.1103/PhysRevLett.121.086602},
  DOI = {10.1103/physrevlett.121.086602},
  number = {8},
  journal = {Physical Review Letters},
  publisher = {American Physical Society (APS)},
  author = {Go,  Dongwook and Jo,  Daegeun and Kim,  Changyoung and Lee,  Hyun-Woo},
  year = {2018},
  month = aug 
}

@article{Go2024,
  title = {First-principles calculation of orbital Hall effect by Wannier interpolation: Role of orbital dependence of the anomalous position},
  volume = {109},
  ISSN = {2469-9969},
  url = {http://dx.doi.org/10.1103/PhysRevB.109.174435},
  DOI = {10.1103/physrevb.109.174435},
  number = {17},
  journal = {Physical Review B},
  publisher = {American Physical Society (APS)},
  author = {Go,  Dongwook and Lee,  Hyun-Woo and Oppeneer,  Peter M. and Bl\"{u}gel,  Stefan and Mokrousov,  Yuriy},
  year = {2024},
  month = may 
}

@article{Choi2023,
  title = {Observation of the orbital Hall effect in a light metal Ti},
  volume = {619},
  ISSN = {1476-4687},
  url = {http://dx.doi.org/10.1038/s41586-023-06101-9},
  DOI = {10.1038/s41586-023-06101-9},
  number = {7968},
  journal = {Nature},
  publisher = {Springer Science and Business Media LLC},
  author = {Choi,  Young-Gwan and Jo,  Daegeun and Ko,  Kyung-Hun and Go,  Dongwook and Kim,  Kyung-Han and Park,  Hee Gyum and Kim,  Changyoung and Min,  Byoung-Chul and Choi,  Gyung-Min and Lee,  Hyun-Woo},
  year = {2023},
  month = jul,
  pages = {52–56}
}

@article{Lyalin2023,
  title = {Magneto-Optical Detection of the Orbital Hall Effect in Chromium},
  volume = {131},
  ISSN = {1079-7114},
  url = {http://dx.doi.org/10.1103/PhysRevLett.131.156702},
  DOI = {10.1103/physrevlett.131.156702},
  number = {15},
  journal = {Physical Review Letters},
  publisher = {American Physical Society (APS)},
  author = {Lyalin,  Igor and Alikhah,  Sanaz and Berritta,  Marco and Oppeneer,  Peter M. and Kawakami,  Roland K.},
  year = {2023},
  month = oct 
}

@article{Kumar2023,
  title = {Ultrafast THz probing of nonlocal orbital current in transverse multilayer metallic heterostructures},
  volume = {14},
  ISSN = {2041-1723},
  url = {http://dx.doi.org/10.1038/s41467-023-43956-y},
  DOI = {10.1038/s41467-023-43956-y},
  number = {1},
  journal = {Nature Communications},
  publisher = {Springer Science and Business Media LLC},
  author = {Kumar,  Sandeep and Kumar,  Sunil},
  year = {2023},
  month = dec 
}

@article{Wang2023,
  title = {Inverse orbital Hall effect and orbitronic terahertz emission observed in the materials with weak spin-orbit coupling},
  volume = {8},
  ISSN = {2397-4648},
  url = {http://dx.doi.org/10.1038/s41535-023-00559-6},
  DOI = {10.1038/s41535-023-00559-6},
  number = {1},
  journal = {npj Quantum Materials},
  publisher = {Springer Science and Business Media LLC},
  author = {Wang,  Ping and Feng,  Zheng and Yang,  Yuhe and Zhang,  Delin and Liu,  Quancheng and Xu,  Zedong and Jia,  Zhiyan and Wu,  Yong and Yu,  Guoqiang and Xu,  Xiaoguang and Jiang,  Yong},
  year = {2023},
  month = may 
}

@article{Krishnia2023,
  title = {Large Interfacial Rashba Interaction Generating Strong Spin–Orbit Torques in Atomically Thin Metallic Heterostructures},
  volume = {23},
  ISSN = {1530-6992},
  url = {http://dx.doi.org/10.1021/acs.nanolett.2c05091},
  DOI = {10.1021/acs.nanolett.2c05091},
  number = {15},
  journal = {Nano Letters},
  publisher = {American Chemical Society (ACS)},
  author = {Krishnia,  Sachin and Sassi,  Yanis and Ajejas,  Fernando and Sebe,  Nicolas and Reyren,  Nicolas and Collin,  Sophie and Denneulin,  Thibaud and Kovács,  András and Dunin-Borkowski,  Rafal E. and Fert,  Albert and George,  Jean-Marie and Cros,  Vincent and Jaffrès,  Henri},
  year = {2023},
  month = jul,
  pages = {6785–6791}
}

@article{Sala2022,
  title = {Giant orbital Hall effect and orbital-to-spin conversion in $3d$, $5d$, and $4f$ metallic heterostructures},
  author = {Sala, Giacomo and Gambardella, Pietro},
  journal = {Phys. Rev. Res.},
  volume = {4},
  issue = {3},
  pages = {033037},
  numpages = {14},
  year = {2022},
  month = {Jul},
  publisher = {American Physical Society},
  doi = {10.1103/PhysRevResearch.4.033037},
  url = {https://link.aps.org/doi/10.1103/PhysRevResearch.4.033037}
}

@article{Ding2020,
  title = {Harnessing Orbital-to-Spin Conversion of Interfacial Orbital Currents for Efficient Spin-Orbit Torques},
  volume = {125},
  ISSN = {1079-7114},
  url = {http://dx.doi.org/10.1103/PhysRevLett.125.177201},
  DOI = {10.1103/physrevlett.125.177201},
  number = {17},
  journal = {Physical Review Letters},
  publisher = {American Physical Society (APS)},
  author = {Ding,  Shilei and Ross,  Andrew and Go,  Dongwook and Baldrati,  Lorenzo and Ren,  Zengyao and Freimuth,  Frank and Becker,  Sven and Kammerbauer,  Fabian and Yang,  Jinbo and Jakob,  Gerhard and Mokrousov,  Yuriy and Kl\"{a}ui,  Mathias},
  year = {2020},
  month = oct 
}

@article{Lee2021,
  title = {Orbital torque in magnetic bilayers},
  volume = {12},
  ISSN = {2041-1723},
  url = {http://dx.doi.org/10.1038/s41467-021-26650-9},
  DOI = {10.1038/s41467-021-26650-9},
  number = {1},
  journal = {Nature Communications},
  publisher = {Springer Science and Business Media LLC},
  author = {Lee,  Dongjoon and Go,  Dongwook and Park,  Hyeon-Jong and Jeong,  Wonmin and Ko,  Hye-Won and Yun,  Deokhyun and Jo,  Daegeun and Lee,  Soogil and Go,  Gyungchoon and Oh,  Jung Hyun and Kim,  Kab-Jin and Park,  Byong-Guk and Min,  Byoung-Chul and Koo,  Hyun Cheol and Lee,  Hyun-Woo and Lee,  OukJae and Lee,  Kyung-Jin},
  year = {2021},
  month = nov 
}

@article{Lee2021b,
  title = {Efficient conversion of orbital Hall current to spin current for spin-orbit torque switching},
  volume = {4},
  ISSN = {2399-3650},
  url = {http://dx.doi.org/10.1038/s42005-021-00737-7},
  DOI = {10.1038/s42005-021-00737-7},
  number = {1},
  journal = {Communications Physics},
  publisher = {Springer Science and Business Media LLC},
  author = {Lee,  Soogil and Kang,  Min-Gu and Go,  Dongwook and Kim,  Dohyoung and Kang,  Jun-Ho and Lee,  Taekhyeon and Lee,  Geun-Hee and Kang,  Jaimin and Lee,  Nyun Jong and Mokrousov,  Yuriy and Kim,  Sanghoon and Kim,  Kab-Jin and Lee,  Kyung-Jin and Park,  Byong-Guk},
  year = {2021},
  month = nov 
}

@article{Gupta2025, title={Harnessing orbital Hall effect in spin-orbit torque MRAM}, volume={16}, ISSN={2041-1723}, url={http://dx.doi.org/10.1038/s41467-024-55437-x}, DOI={10.1038/s41467-024-55437-x}, number={1}, journal={Nature Communications}, publisher={Springer Science and Business Media LLC}, author={Gupta, Rahul and Bouard, Chloé and Kammerbauer, Fabian and Ledesma-Martin, J. Omar and Bose, Arnab and Kononenko, Iryna and Martin, Sylvain and Usé, Perrine and Jakob, Gerhard and Drouard, Marc and Kläui, Mathias}, year={2025}, month=jan }

@ARTICLE{Salemi2022,
  title     = "First-principles theory of intrinsic spin and orbital Hall and
               Nernst effects in metallic monoatomic crystals",
  author    = "Salemi, Leandro and Oppeneer, Peter M",
  journal   = "Phys. Rev. Mater.",
  publisher = "American Physical Society (APS)",
  volume    =  6,
  number    =  9,
  month     =  sep,
  year      =  2022,
  copyright = "https://creativecommons.org/licenses/by/4.0/",
  language  = "en"
}

@inproceedings{Li2023,
  author={Li, Kai-Shin and Shieh, Jia-Min and Chen, Yi-Ju and Hsu, Cho-Lun and Shen, Chang-Hong and Hou, Tuo-Hung and Lin, Chia-Ping and Lai, Chih-Huang and Tang, Denny D. and Yuan-Chen Sun, Jack},
  title={First BEOL-compatible, 10 ns-fast, and Durable 55 nm Top-pSOT-MRAM with High TMR ($>$ 130\%)}, 
  year={2023},
  volume={},
  number={},
  pages={1-4},
  booktitle={2023 International Electron Devices Meeting (IEDM)},
  doi={10.1109/IEDM45741.2023.10413685}}

@article{yang2008,
	title = {Giant spin-accumulation signal and pure spin-current-induced reversible magnetization switching},
	volume = {4},
	rights = {2008 Nature Publishing Group},
	issn = {1745-2481},
	url = {https://www.nature.com/articles/nphys1095},
	doi = {10.1038/nphys1095},
	pages = {851--854},
	number = {11},
	journaltitle = {Nature Physics},
	shortjournal = {Nature Phys},
	author = {Yang, Tao and Kimura, Takashi and Otani, Yoshichika},
	urldate = {2023-03-22},
	date = {2008-11},
        year = {2008},
	langid = {english},
	note = {Number: 11
Publisher: Nature Publishing Group}
}

@article{Valenzuela2006,
  title = {Direct electronic measurement of the spin Hall effect},
  volume = {442},
  ISSN = {1476-4687},
  url = {http://dx.doi.org/10.1038/nature04937},
  DOI = {10.1038/nature04937},
  number = {7099},
  journal = {Nature},
  publisher = {Springer Science and Business Media LLC},
  author = {Valenzuela,  S. O. and Tinkham,  M.},
  year = {2006},
  month = jul,
  pages = {176–179}
}

@article{Sun2009,
  title = {A three-terminal spin-torque-driven magnetic switch},
  volume = {95},
  ISSN = {1077-3118},
  url = {http://dx.doi.org/10.1063/1.3216851},
  DOI = {10.1063/1.3216851},
  number = {8},
  journal = {Applied Physics Letters},
  publisher = {AIP Publishing},
  author = {Sun,  J. Z. and Gaidis,  M. C. and O’Sullivan,  E. J. and Joseph,  E. A. and Hu,  G. and Abraham,  D. W. and Nowak,  J. J. and Trouilloud,  P. L. and Lu,  Yu and Brown,  S. L. and Worledge,  D. C. and Gallagher,  W. J.},
  year = {2009},
  month = aug 
}

@article{Garello2013,
  title = {Symmetry and magnitude of spin–orbit torques in ferromagnetic heterostructures},
  volume = {8},
  ISSN = {1748-3395},
  url = {http://dx.doi.org/10.1038/nnano.2013.145},
  DOI = {10.1038/nnano.2013.145},
  number = {8},
  journal = {Nature Nanotechnology},
  publisher = {Springer Science and Business Media LLC},
  author = {Garello,  Kevin and Miron,  Ioan Mihai and Avci,  Can Onur and Freimuth,  Frank and Mokrousov,  Yuriy and Bl\"{u}gel,  Stefan and Auffret,  Stéphane and Boulle,  Olivier and Gaudin,  Gilles and Gambardella,  Pietro},
  year = {2013},
  month = jul,
  pages = {587–593}
}

@article{Dutta2022,
  title = {Observation of nonlocal orbital transport and sign reversal of dampinglike torque in Nb/Ni and Ta/Ni bilayers},
  volume = {106},
  ISSN = {2469-9969},
  url = {http://dx.doi.org/10.1103/PhysRevB.106.184406},
  DOI = {10.1103/physrevb.106.184406},
  number = {18},
  journal = {Physical Review B},
  publisher = {American Physical Society (APS)},
  author = {Dutta,  Sutapa and Tulapurkar,  Ashwin A.},
  year = {2022},
  month = nov 
}

@article{Pai2012,
  doi = {10.1063/1.4753947},
  url = {https://doi.org/10.1063/1.4753947},
  year = {2012},
  month = sep,
  publisher = {{AIP} Publishing},
  volume = {101},
  number = {12},
  pages = {122404},
  author = {Chi-Feng Pai and Luqiao Liu and Y. Li and H. W. Tseng and D. C. Ralph and R. A. Buhrman},
  title = {Spin transfer torque devices utilizing the giant spin Hall effect of tungsten},
  journal = {Applied Physics Letters}
}

@article{Manchon2019,
  title = {Current-induced spin-orbit torques in ferromagnetic and antiferromagnetic systems},
  volume = {91},
  ISSN = {1539-0756},
  url = {http://dx.doi.org/10.1103/RevModPhys.91.035004},
  DOI = {10.1103/revmodphys.91.035004},
  number = {3},
  journal = {Reviews of Modern Physics},
  publisher = {American Physical Society (APS)},
  author = {Manchon,  A. and Železný,  J. and Miron,  I. M. and Jungwirth,  T. and Sinova,  J. and Thiaville,  A. and Garello,  K. and Gambardella,  P.},
  year = {2019},
  month = sep 
}

@article{Liu2011,
  title = {Spin-Torque Ferromagnetic Resonance Induced by the Spin Hall Effect},
  volume = {106},
  ISSN = {1079-7114},
  url = {http://dx.doi.org/10.1103/PhysRevLett.106.036601},
  DOI = {10.1103/physrevlett.106.036601},
  number = {3},
  journal = {Physical Review Letters},
  publisher = {American Physical Society (APS)},
  author = {Liu,  Luqiao and Moriyama,  Takahiro and Ralph,  D. C. and Buhrman,  R. A.},
  year = {2011},
  month = jan 
}

@article{Vudya2021,
  title = {Optimization of Tungsten $\beta$-Phase Window for Spin-Orbit-Torque Magnetic Random-Access Memory},
  volume = {16},
  ISSN = {2331-7019},
  url = {http://dx.doi.org/10.1103/PhysRevApplied.16.064009},
  DOI = {10.1103/physrevapplied.16.064009},
  number = {6},
  journal = {Physical Review Applied},
  publisher = {American Physical Society (APS)},
  author = {Vudya Sethu,  Kiran Kumar and Ghosh,  Sambit and Couet,  Sebastien and Swerts,  Johan and Sorée,  Bart and De Boeck,  Jo and Kar,  Gouri Sankar and Garello,  Kevin},
  year = {2021},
  month = dec 
}

@article{McHugh2020,
  title = {Impact of impurities on the spin Hall conductivity in $\beta$-W},
  volume = {4},
  ISSN = {2475-9953},
  url = {http://dx.doi.org/10.1103/PhysRevMaterials.4.094404},
  DOI = {10.1103/physrevmaterials.4.094404},
  number = {9},
  journal = {Physical Review Materials},
  publisher = {American Physical Society (APS)},
  author = {McHugh,  Oliver L. W. and Goh,  Wen Fong and Gradhand,  Martin and Stewart,  Derek A.},
  year = {2020},
  month = sep 
}

@article{Garello2014,
  title = {Ultrafast magnetization switching by spin-orbit torques},
  volume = {105},
  ISSN = {1077-3118},
  url = {http://dx.doi.org/10.1063/1.4902443},
  DOI = {10.1063/1.4902443},
  number = {21},
  journal = {Applied Physics Letters},
  publisher = {AIP Publishing},
  author = {Garello,  Kevin and Avci,  Can Onur and Miron,  Ioan Mihai and Baumgartner,  Manuel and Ghosh,  Abhijit and Auffret,  Stéphane and Boulle,  Olivier and Gaudin,  Gilles and Gambardella,  Pietro},
  year = {2014},
  month = nov 
}

@article{Huang2025,
  title = {Top Spin-Orbit-Torque Switching of Magnetic Tunnel Junction With In-Situ Efficiency Quantification},
  volume = {46},
  ISSN = {1558-0563},
  url = {http://dx.doi.org/10.1109/LED.2025.3582335},
  DOI = {10.1109/led.2025.3582335},
  number = {8},
  journal = {IEEE Electron Device Letters},
  publisher = {Institute of Electrical and Electronics Engineers (IEEE)},
  author = {Huang,  Yan and Zhang,  Kun and Xiao,  Chen and Yang,  Qing and Xu,  Shijie and Cai,  Wenlong and Zhu,  Daoqian and Yang,  Jiakai and Zhao,  Weisheng},
  year = {2025},
  month = aug,
  pages = {1409–1412}
}

@article{shao_roadmap_2021,
    title = {Roadmap of spin–orbit torques},
    volume = {57},
    issn = {0018-9464},
    number = {7},
    journal = {IEEE Transactions on Magnetics},
    author = {Shao, Qiming and Li, Peng and Liu, Luqiao and Yang, Hyunsoo and Fukami, Shunsuke and Razavi, Armin and Wu, Hao and Wang, Kang and Freimuth, Frank and Mokrousov, Yuriy},
    year = {2021},
    note = {Publisher: IEEE},
    pages = {1--39},
}

@article{Gao2025,
  title = {Nonlocal electrical detection of reciprocal orbital Edelstein effect},
  volume = {16},
  ISSN = {2041-1723},
  url = {http://dx.doi.org/10.1038/s41467-025-61602-7},
  DOI = {10.1038/s41467-025-61602-7},
  number = {1},
  journal = {Nature Communications},
  publisher = {Springer Science and Business Media LLC},
  author = {Gao,  Weiguang and Liao,  Liyang and Isshiki,  Hironari and Budai,  Nico and Kim,  Junyeon and Lee,  Hyun-Woo and Lee,  Kyung-Jin and Go,  Dongwook and Mokrousov,  Yuriy and Miwa,  Shinji and Otani,  Yoshichika},
  year = {2025},
  month = jul 
}

@article{Dyakonov1971,
  title = {Current-induced spin orientation of electrons in semiconductors},
  volume = {35},
  ISSN = {0375-9601},
  url = {http://dx.doi.org/10.1016/0375-9601(71)90196-4},
  DOI = {10.1016/0375-9601(71)90196-4},
  number = {6},
  journal = {Physics Letters A},
  publisher = {Elsevier BV},
  author = {Dyakonov,  M.I. and Perel,  V.I.},
  year = {1971},
  month = jul,
  pages = {459–460}
}

@article{Sinova2015,
  title = {Spin Hall effects},
  volume = {87},
  ISSN = {1539-0756},
  url = {http://dx.doi.org/10.1103/RevModPhys.87.1213},
  DOI = {10.1103/revmodphys.87.1213},
  number = {4},
  journal = {Reviews of Modern Physics},
  publisher = {American Physical Society (APS)},
  author = {Sinova,  Jairo and Valenzuela,  Sergio O. and Wunderlich,  J. and Back,  C. H. and Jungwirth,  T.},
  year = {2015},
  month = oct,
  pages = {1213–1260}
}

@inproceedings{VanBeek2023,
  title = {Scaling the SOT track – A path towards maximizing efficiency in SOT-MRAM},
  url = {http://dx.doi.org/10.1109/IEDM45741.2023.10413749},
  DOI = {10.1109/iedm45741.2023.10413749},
  booktitle = {2023 International Electron Devices Meeting (IEDM)},
  publisher = {IEEE},
  author = {Van Beek,  S. and Cai,  K. and Yasin,  F. and Hody,  H. and Talmelli,  G. and Nguyen,  V.D. and Vergel,  N. Franchina and Palomino,  A. and Trovato,  A. and Wostyn,  K. and Rao,  S. and Kar,  G.S. and Couet,  S.},
  year = {2023},
  month = dec,
  pages = {1–4}
}

@INPROCEEDINGS{Yasin2024,
  author={Yasin, F. and Palomino, A. and Kumar, A. and Pica, V. and Van Beek, S. and Talmelli, G. and Nguyen, V.D. and Cosemans, S. and Crotti, D. and Wostyn, K. and Kar, G. S. and Couet, S.},
  booktitle={2024 IEEE Symposium on VLSI Technology and Circuits (VLSI Technology and Circuits)}, 
  title={Extremely Scaled Perpendicular SOT-MRAM Array Integration on 300mm Wafer}, 
  year={2024},
  volume={},
  number={},
  pages={1-2},
  doi={10.1109/VLSITechnologyandCir46783.2024.10631340}}

@article{Rothschild2022,
  title = {Generation of spin currents by the orbital Hall effect in Cu and Al and their measurement by a Ferris-wheel ferromagnetic resonance technique at the wafer level},
  volume = {106},
  ISSN = {2469-9969},
  url = {http://dx.doi.org/10.1103/PhysRevB.106.144415},
  DOI = {10.1103/physrevb.106.144415},
  number = {14},
  journal = {Physical Review B},
  publisher = {American Physical Society (APS)},
  author = {Rothschild,  Amit and Am-Shalom,  Nadav and Bernstein,  Nirel and Meron,  Ma’yan and David,  Tal and Assouline,  Benjamin and Frohlich,  Elichai and Xiao,  Jiewen and Yan,  Binghai and Capua,  Amir},
  year = {2022},
  month = oct 
}

@article{Go2020,
  title = {Theory of current-induced angular momentum transfer dynamics in spin-orbit coupled systems},
  volume = {2},
  ISSN = {2643-1564},
  url = {http://dx.doi.org/10.1103/PhysRevResearch.2.033401},
  DOI = {10.1103/physrevresearch.2.033401},
  number = {3},
  journal = {Physical Review Research},
  publisher = {American Physical Society (APS)},
  author = {Go,  Dongwook and Freimuth,  Frank and Hanke,  Jan-Philipp and Xue,  Fei and Gomonay,  Olena and Lee,  Kyung-Jin and Bl\"{u}gel,  Stefan and Haney,  Paul M. and Lee,  Hyun-Woo and Mokrousov,  Yuriy},
  year = {2020},
  month = sep 
}

@article{Amin2019,
  title = {Intrinsic spin currents in ferromagnets},
  volume = {99},
  ISSN = {2469-9969},
  url = {http://dx.doi.org/10.1103/PhysRevB.99.220405},
  DOI = {10.1103/physrevb.99.220405},
  number = {22},
  journal = {Physical Review B},
  publisher = {American Physical Society (APS)},
  author = {Amin,  V. P. and Li,  Junwen and Stiles,  M. D. and Haney,  P. M.},
  year = {2019},
  month = jun 
}

@article{Molas2021,
  title = {Advances in Emerging Memory Technologies: From Data Storage to Artificial Intelligence},
  volume = {11},
  ISSN = {2076-3417},
  url = {http://dx.doi.org/10.3390/app112311254},
  DOI = {10.3390/app112311254},
  number = {23},
  journal = {Applied Sciences},
  publisher = {MDPI AG},
  author = {Molas,  Gabriel and Nowak,  Etienne},
  year = {2021},
  month = nov,
  pages = {11254}
}

@INPROCEEDINGS{Liu2017,
  author={Swerts, J. and Liu, E. and Couet, S. and Mertens, S. and Rao, S. and Kim, W. and Garello, K. and Souriau, L. and Kundu, S. and Crotti, D. and Yasin, F. and Jossart, N. and Sakhare, S. and Devolder, T. and Van Beek, S. and O'Sullivan, B. and Van Elshocht, S. and Furnemont, A. and Kar, G. S.},
  booktitle={2017 IEEE International Electron Devices Meeting (IEDM)}, 
  title={Solving the BEOL compatibility challenge of top-pinned magnetic tunnel junction stacks}, 
  year={2017},
  volume={},
  number={},
  pages={38.6.1-38.6.4},
  doi={10.1109/IEDM.2017.8268518}}

@misc{Valet2024,
  doi = {10.48550/ARXIV.2507.06771},
  url = {https://arxiv.org/abs/2507.06771},
  author = {Valet,  T. and Jaffres,  H. and Cros,  V. and Raimondi,  R.},
  title = {Quantum Kinetic Anatomy of Electron Angular Momenta Edge Accumulation},
  publisher = {arXiv},
  year = {2025},
  copyright = {Creative Commons Attribution 4.0 International}
}

@misc{NLpatent,
    author = {Viala, Bernard and Garello, Kevin},
    title = {Orbital hall effect magnetic device and method for manufacturing such a device},
    year = {2025},
    howpublished = {US Patent US20250338505A1}
}

@article{Berger2018,
  title = {Determination of the spin Hall effect and the spin diffusion length of Pt from self-consistent fitting of damping enhancement and inverse spin-orbit torque measurements},
  volume = {98},
  ISSN = {2469-9969},
  url = {http://dx.doi.org/10.1103/PhysRevB.98.024402},
  DOI = {10.1103/physrevb.98.024402},
  number = {2},
  journal = {Physical Review B},
  publisher = {American Physical Society (APS)},
  author = {Berger,  Andrew J. and Edwards,  Eric R. J. and Nembach,  Hans T. and Karis,  Olof and Weiler,  Mathias and Silva,  T. J.},
  year = {2018},
  month = July 
}

@article{Dolui2017,
  title = {Spin-memory loss due to spin-orbit coupling at ferromagnet/heavy-metal interfaces: Ab initio spin-density matrix approach},
  volume = {96},
  ISSN = {2469-9969},
  url = {http://dx.doi.org/10.1103/PhysRevB.96.220403},
  DOI = {10.1103/physrevb.96.220403},
  number = {22},
  journal = {Physical Review B},
  publisher = {American Physical Society (APS)},
  author = {Dolui,  Kapildeb and Nikolić,  Branislav K.},
  year = {2017},
  month = Dec 
}

@article{Han2021,
  title = {The thickness dependence of the field-like spin–orbit torque in heavy metal/CoFeB/MgO heterostructures},
  volume = {130},
  ISSN = {1089-7550},
  url = {http://dx.doi.org/10.1063/5.0068373},
  DOI = {10.1063/5.0068373},
  number = {21},
  journal = {Journal of Applied Physics},
  publisher = {AIP Publishing},
  author = {Han,  Bo and Zhang,  Bo and Sun,  Shuling and Wang,  Bo and Guo,  Yonghai and Cao,  Jiangwei},
  year = {2021},
  month = Dec 
}

@article{Berger1972,
  title = {Models for contacts to planar devices},
  volume = {15},
  ISSN = {0038-1101},
  url = {http://dx.doi.org/10.1016/0038-1101(72)90048-2},
  DOI = {10.1016/0038-1101(72)90048-2},
  number = {2},
  journal = {Solid-State Electronics},
  publisher = {Elsevier BV},
  author = {Berger,  H.H.},
  year = {1972},
  month = Feb,
  pages = {145–158}
}

\end{document}